\documentclass[conference]{IEEEtran}
\IEEEoverridecommandlockouts
\usepackage{cite}
\usepackage{amsmath,amssymb,amsfonts}
\usepackage{amsthm}
\usepackage{algorithmic}
\usepackage[caption=false,font=normalsize,labelfont=sf,textfont=sf]{subfig}
\usepackage{graphicx}
\usepackage{textcomp}
\usepackage{stfloats}
\usepackage{url}
\usepackage{verbatim}
\usepackage{cite}
\usepackage[dvipsnames]{xcolor}
\usepackage{bbm}
\usepackage{enumitem}


\def\BibTeX{{\rm B\kern-.05em{\sc i\kern-.025em b}\kern-.08em
    T\kern-.1667em\lower.7ex\hbox{E}\kern-.125emX}}
\input{mysymbol.sty}
\begin{document}

\title{Biased Backpressure Routing for Multihop Wireless Networks with Heterogeneous Interfaces\\
\thanks{Research was sponsored by the U.S. Army Combat Capabilities Development Command Army Research Laboratory (DEVCOM ARL) Army Research Office and was accomplished under Cooperative Agreement Numbers W911NF-24-2-0008 and W911NF-26-2-A128. The views and conclusions contained in this document are those of the authors and should not be interpreted as representing the official policies, either expressed or implied, of the Army Research Office or the U.S. Government. The U.S. Government is authorized to reproduce and distribute reprints for Government purposes notwithstanding any copyright notation herein.}
\thanks{Emails: \{\texttt{yujun.ming}, \texttt{zhongyuan.zhao}, \texttt{segarra}\}@rice.edu, \{\texttt{justin.h.kong2}, \texttt{terrence.j.moore}, \texttt{fikadu.t.dagefu}, \texttt{ananthram.swami}, \texttt{kevin.s.chan}\}.civ@army.mil }
}

\author{
Yujun Ming$^{\star}$, Zhongyuan Zhao$^{\star}$, Fikadu Dagefu$^{\ddag}$, Justin Kong$^{\ddag}$, Terrence Moore$^{\ddag}$,\\ 
Kevin Chan$^{\ddag}$, Ananthram Swami$^{\ddag}$,
and {Santiago Segarra}$^{\star}$\\
\textit{$^\star$Rice University}, Houston, TX, USA\\
\textit{$^\ddag$DEVCOM Army Research Laboratory}, Adelphi, MD, USA
}

\maketitle

\begin{abstract}
Heterogeneous-interface multihop wireless networks (Het-MuNets) are emerging as a promising paradigm for tactical networks and for infrastructure-light applications such as vehicular communications, wireless backhaul, and non-terrestrial connectivity.
To exploit the diverse profiles of heterogeneous communication technologies in penetration, interference, and bandwidth, packet-to-interface assignment must be determined on a \emph{per-hop} basis, making routing and scheduling highly complex.
In this work, we develop a unified framework for joint packet routing, link scheduling, and interface assignment in Het-MuNets with multiple concurrent flows. 
By modeling packet-to-interface assignment as transmission between virtual subnodes, we transform interface assignment into intra-device virtual routing, which is solved jointly with physical routing and scheduling under a unified multi-layer shortest path-biased Backpressure (SP-BP) scheme.
Numerical results demonstrate that the proposed framework outperforms SP-BP operating on other baseline graph models and non-backpressure routing schemes in goodput, latency, and packet delivery rate.
\end{abstract}


\section{Introduction}
\label{sec:intro}
Heterogeneous-interface multihop wireless networks (Het-MuNets) offer a powerful, infrastructure-light paradigm for applications like tactical networking, robotic swarms, wireless backhaul, and non-terrestrial coverage~\cite{al2015channel,akyildiz20206g,kott2016internet,mahmud2021software}.
By combining diverse communication technologies with complementary physical capabilities, Het-MuNets can simultaneously achieve multiple conflicting objectives that are unsupported by a single technology~\cite{sylla2022multi}. 
For example, a network can achieve high throughput and covertness with mmWave or visible light communications (VLC), while maintaining robust connectivity in long-range and mobile scenarios using the high penetration of sub-GHz bands~\cite{10571922}.
It can also balance quality-of-service, energy efficiency and cost for different traffic demands by switching between 5G New Radio (NR) and co-band protocols (e.g., Wi-Fi and Bluetooth). 
Although conceptually adjacent to HetNets, multi-RAT systems~\cite{9896125,akyildiz20206g},  Multi-Bearer Networks (MBNs)~\cite{mahmud2021software}, and multi-radio multi-channel (MR-MC) networks~\cite{li2012mrmc,al2015channel}, Het-MuNets emphasize multi-hop packet routing across diverse communication technologies rather than cellular architecture or link-level waveform switching.
The primary challenge in Het-MuNets is that interface assignment dictates the topological inputs for routing and scheduling. 
Joint optimization thus becomes significantly harder with this added degree of freedom, especially under heterogeneous device profiles.

Classical multihop routing protocols, such as AODV~\cite{749281}, establish a single path per flow using static additive metrics (e.g., hop count), potentially congesting bottleneck links shared by concurrent flows.
Although AODVv2~\cite{perkins2016ad} introduces multi-interface support, interface assignments remain locked after route discovery and cannot adapt to dynamic congestion.
To reduce the signaling overhead of global topology tracking, reinforcement learning (RL) has been adopted for on-demand routing in Het-MuNets, including Q-learning for covert routing~\cite{kong2024decentralized} and a shared deep Q-network that jointly selects the next-hop, interface, and sub-band during route establishment~\cite{kim2025deep}. 
However, these single-path routing approaches are generally vulnerable to link failures, multi-flow resource contention, and real-time congestion dynamics.

Backpressure (BP) routing addresses these drawbacks through joint routing and scheduling based on queue differentials (pressure) and supports fully distributed operation with provable queue stability established via Lyapunov drift-plus-penalty (DPP) analysis~\cite{neely2005dynamic,neely2010stochastic}.
By incorporating distance-based biases into pressure calculations, shortest path-biased BP (SP-BP)~\cite{neely2005dynamic,zhao2024tmlcn} mitigates the slow startup and random-walk behavior of classical BP, thus significantly reducing the latency under light-to-moderate loads~\cite{ji2012delay}.
Since the biases are queue-agnostic and updated infrequently to capture topological drift, the signaling overhead of SP-BP remains low~\cite{zhao2024tmlcn}. 
However, extending homogeneous SP-BP to Het-MuNets remains an open and non-trivial challenge.
Closely related schemes include BP for MR-MC and MIMO networks that assume parallel homogeneous interfaces~\cite{li2012mrmc,zhao2026ton}.

\begin{figure*}[!t]
    \centering
    \includegraphics[width=0.85\linewidth]{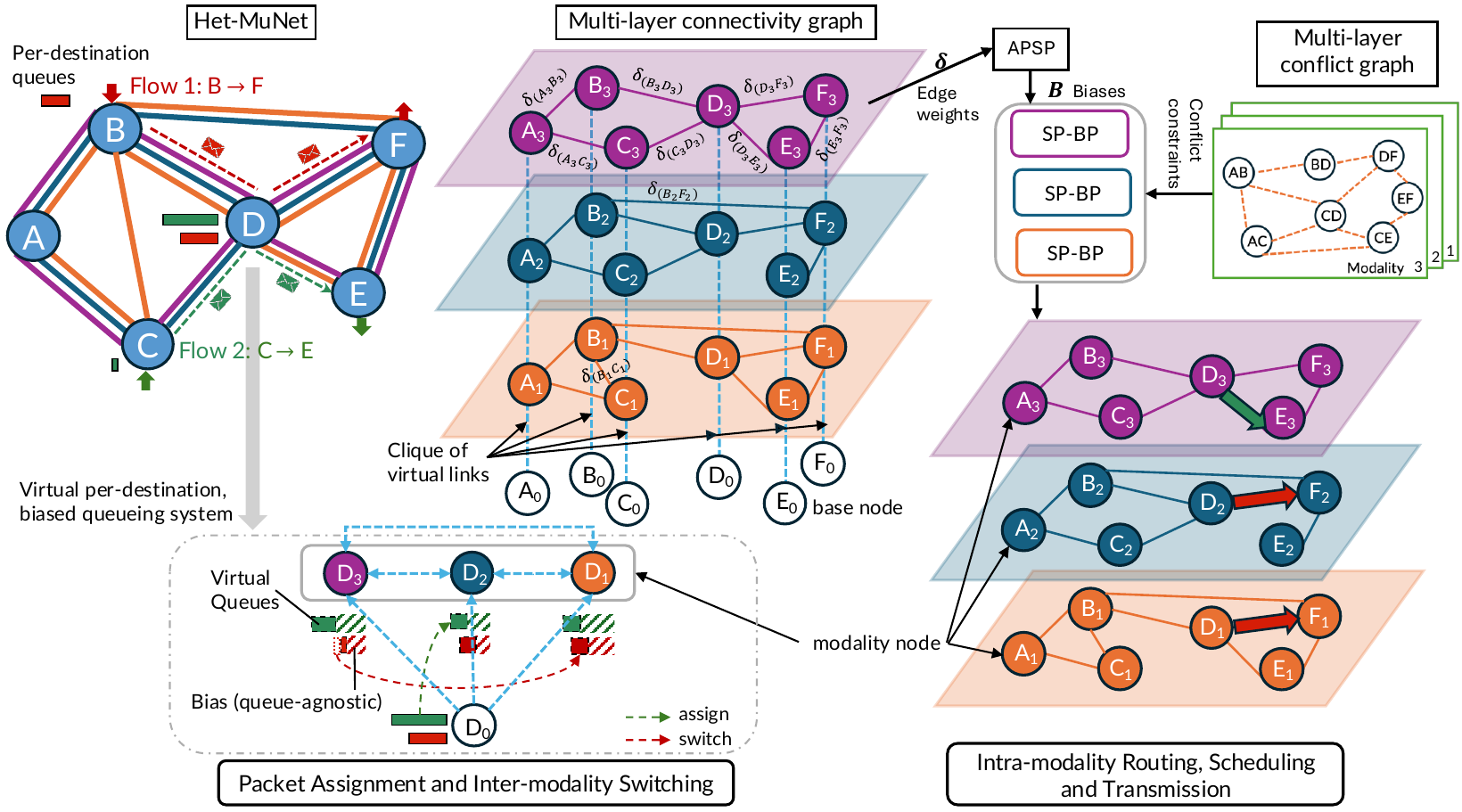}
    \vspace{-0.4em}
    \caption{System diagram of multi-layer BP routing for Het-MuNets: multi-layer graph modeling, queueing system design, and operations.}
    \vspace{-1.2em}
\label{fig:system}
\end{figure*}

In this work, we propose multi-layer SP-BP as a multi-flow multi-path routing scheme for Het-MuNets.
The key innovation is a multi-layer graph model that represents heterogeneous air-interfaces on a device as subnodes connected by virtual links, and the sub-network of each communication technology as a graph layer (Fig.~\ref{fig:system}).
This approach transforms cross-interface packet assignment into intra-device virtual routing.
As a result, interface assignment, packet routing, and link scheduling can be optimized jointly under a unified Backpressure-MaxWeight framework.
Furthermore, we introduce virtual link weights and a coupled actual-virtual queueing system to eliminate the intra-device packet looping in the existing graph and queueing-system expansions~\cite{li2012mrmc}.

\noindent\textbf{Contributions:} Our contributions are as follows:
\begin{itemize}[leftmargin=1em,labelsep=0.2em,itemsep=0pt,topsep=1pt]
    \item We introduce a multi-layer graph model and a system of actual-virtual queues for networking in Het-MuNets, enabling fully distributed SP-BP to jointly optimize interface assignment, packet routing, and link scheduling.
    \item We develop a multi-layer SP-BP scheme that solves virtual and physical routing in two stages coupled by virtual queueing states. The intra-device virtual routing adopts a minimum-cost flow (MCF) formulation that can be solved exactly in polynomial time, while the physical routing and scheduling rely on SP-BP for individual interface layers.
    \item Through numerical experiments under mixed streaming and bursty traffic patterns and various network sizes, we show that the proposed multi-layer routing algorithm reduces latency and improves packet delivery and goodput compared with both BP and non-BP baselines.
\end{itemize}
\noindent
{\bf Notation:} 
$ |\cdot| $ represents the cardinality of a set.
$ \mathbb{E}(\cdot) $ stands for expectation.
Upright bold lower-case symbol, e.g., $\bbz$, denotes a column vector, and $\bbz_i$ denotes the $i$-th element of vector $\bbz$. 
Upright bold upper-case symbol $\bbZ$ denotes a matrix, and $\bbZ_{i,j}$ for its element at row $i$ and column $j$. Calligraphic upper-case symbol denotes a set, e.g., $\ccalG$ for a graph.

\section{System Model}\label{sec:system}
A Het-MuNet can be modeled as a directed multigraph $\ccalG^n = (\ccalV^0, \ccalE^p)$, where a node $i\in\ccalV^0$ represents a device, and an edge $e^p=(i_m,j_m)\in\ccalE^p$ indicates that device $i$ can transmit data to device $j$ directly via communication modality $m \in \ccalM$, where $\ccalM$ is the set of all modalities in the network.
There can be multiple edges from $i$ to $j$.
We define \emph{communication modalities} as wireless technologies with potentially mutually exclusive activations, such as 5G NR, Wi-Fi, Bluetooth, UHF radio, or VLC.
We denote the set of available modalities on device $i$ as $\ccalM_i\subseteq\ccalM$.

To unify modality assignment, packet routing, and link scheduling under the same decision framework, we  convert the multigraph $\ccalG^n$ into a multi-layer graph model $\ccalG^m=({\ccalV},\ccalE)$ where $\ccalE=\ccalE^p \cup \ccalE^v$, as illustrated in Fig.~\ref{fig:system}.
We define the set $\ccalV^0$ (reused from $\ccalG^n$ as it includes all devices) as subnodes in the base layer and the set $\ccalV^m$ as subnodes in the layer of modality $m$.
The vertex set is ${\ccalV}=\bigcup_{m\in\ccalM^+}\ccalV^m$, where $\ccalM^+=\{0\}\cup\ccalM$. 
On device $i$, the base node $i_0\in\ccalV^0$ serves as the injection point of exogenous arrivals, whereas a modality node $i_m\in\ccalV^m$ handles endogenous traffic. 
Any packets destined to device $i$ will be immediately consumed upon arriving at any modality node $i_m$; therefore, $\ccalG^m$ contains a built-in sink layer, which is omitted for simplicity.
A physical modality-link $e^p=(i_m,j_m)\in\ccalE^p$ belongs to layer $m$. 
Within a device $i$, we model packet-modality switching as intra-device virtual links $e^v\in\ccalE^v$, e.g., $e^v=(i_0,i_m)$ for assigning newly arrived packets to modality $m$, and $e^v=(i_m,i_{m'})$ for assigning packets from modality $m$ to $m'$.
Note that base node $i_0$ has only outgoing virtual links, i.e., $(i_m,i_0)\notin\ccalE^v$.

We model resource contention between physical links as a conflict graph $\ccalG^c = (\ccalE^p, \ccalH)$ from $\ccalG^n$. 
Each vertex $e\in\ccalE^p$ corresponds to a modality-link in $\ccalG^n$, while an undirected edge $(e_1,e_2)\in\ccalH$ indicates a conflict between links \(e_1\) and \(e_2\). 
Two types of conflicts are considered: 1) \emph{interface conflicts} from transceiver limitations (e.g., shared RF chains); and 2) \emph{interference conflicts} from simultaneous co-band or nearby transmissions. 
For example, two omni-directional links using modality \(m\) conflict if their incident nodes are within interference range \(\rho_m\).
Graph $\ccalG^c$ can be decomposed into modality layers, $\{\widetilde{\ccalG}^c_m\}_{ m\in\ccalM}$, and cross-layer conflict edges.
Notice that our fully distributed SP-BP does not require global knowledge of $\ccalG^c$ or $\widetilde\ccalG^c_m$; instead, each modality-link only needs to know its conflicting neighbors via channel monitoring and network feedback~\cite{zhao2022twc,zhao2024tmlcn,zhao2026ton}.

We consider a time-slotted orthogonal multiple access system, in which a valid schedule must be a set of non-conflicting links.
We denote the instantaneous rate of a modality-link \(e^p\in\ccalE^p\) in time slot \(t\) by \(\grave{\bbr}_{e^p}(t)\), its long-term rate by \(\bbr_{e^p} = \mathbb{E}[\grave{\bbr}_{e^p}(t)]\), and the long-term rate vector by~\(\bbr = [\bbr_{e^p}\mid e^p \in \ccalE^p]\), all measured in packets per slot. 

We refer to all packets destined to device \(c\!\in\!\ccalV^0\) as commodity \(c\). 
Each device $i$ hosts a separate physical queue $\ccalQ_i^{(c)}$ for each commodity $c$. The queue length of $\ccalQ_i^{(c)}$, denoted as $Q_i^{(c)}\!(t)$, evolves as
\begin{equation}\label{E:queue}
Q_{i}^{(c)}(t\!+\!1)=Q_{i}^{(c)}\!(t)-\!\!\!\!\!\sum_{e^p\in\ccalE_i^{p,+}}\!\!\! \mu_{e^p}^{(c)}\!(t)+\!\!\!\!\!\sum_{e^p\in\ccalE_i^{p,-}}\!\!\!\mu_{e^p}^{(c)}\!(t)+A_{i}^{(c)}\!(t),
\vspace{-0.2em}
\end{equation}
where \(\mu_{e^p}^{(c)}(t)\le\grave{\bbr}_{e^p}(t)\) is the transmitted
packet count, \(A_i^{(c)}(t)\) denotes exogenous arrivals, and
\(\ccalE_i^{p,+}\) and \(\ccalE_i^{p,-}\) are the sets of outgoing and
incoming physical links of device $i$, respectively.

Each modality subnode $i_m$ maintains a virtual queue $q_{i_m}^{(c)}(t)$ counting the physical packets assigned to modality $m$. Before new arrivals are assigned in slot $t$, these queues partition the physical queue: $Q_i^{(c)}(t)=\sum_{m\in\ccalM_i}q_{i_m}^{(c)}(t)$. The base node $i_0$ holds only new arrivals, $q_{i_0}^{(c)}(t)=A_i^{(c)}(t)$, and Stage~V changes only their modality assignments without creating or removing packets.
The \textit{decision variables} for the numbers of packets to be transmitted on virtual link $e^v$ and physical link $e^p$ are $\nu_{e^v}^{(c)}(t)$ and $\mu_{e^p}^{(c)}(t)$, respectively.
The virtual queue of commodity $c$ on modality-node $i_m$ evolves as
\begin{equation}\label{eq:virtual_queue_dynamics}
q_{i_m}^{(c)}(t+1)=q_{i_m}^{(c)}(t)+\Delta\nu_{i_m}^{(c)}(t)+\Delta\mu_{i_m}^{(c)}(t)\;,
\end{equation}
where $\Delta\nu_{i_m}^{(c)}(t)$ and
$\Delta\mu_{i_m}^{(c)}(t)$ denote the virtual and physical net inflows,
respectively:
$$
\Delta\nu_{i_m}^{(c)}(t)=\!\!\!\!\sum_{e\in\ccalE^{v,-}_{i_m}}\!\!\!\nu_{e}^{(c)}(t)-\!\!\!\!\sum_{e\in\ccalE^{v,+}_{i_m}}\!\!\!\nu_{e}^{(c)}(t) \;,
$$
$$ 
\Delta\mu_{i_m}^{(c)}(t)=\!\!\!\!\sum_{e\in\ccalE_{i_m}^{p,-}}\!\!\!\mu_{e}^{(c)}(t)-\!\!\!\!\sum_{e\in\ccalE_{i_m}^{p,+}}\!\!\!\mu_{e}^{(c)}(t)\;.
$$
Here, \(+\) and \(-\) identify outgoing and incoming link sets, while
\(p\) and \(v\) specify the physical and virtual link types.
\section{Biased Backpressure for Het-MuNets}\label{sec:solution}
\begin{figure}[t]
    \centering
    \includegraphics[width=0.95\linewidth]{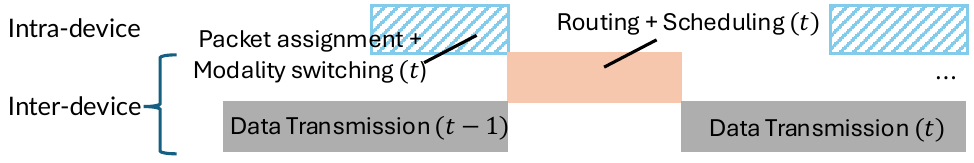}
    \vspace{-0.4em}
    \caption{Operations Timeline in multi-layer SP-BP.}\label{fig:timeline}
    \vspace{-0.2in}
\end{figure}

The proposed multi-layer SP-BP is a fully distributed scheme, where each subnode makes decisions using only information about its own virtual queues and those of its immediate neighbors on $\ccalG^m$, together with the instantaneous rates of its incident links.
As illustrated in Fig.~\ref{fig:timeline}, each slot performs intra-device computation followed by inter-device routing, maximum weight independent set (MWIS) scheduling, and data transmission. 
Data transmission of slot $t$ is pipelined with the subsequent slot $t+1$ intra-device computation.
A brief overview of these stages follows:
\begin{itemize}[leftmargin=1em,labelsep=0.2em,itemsep=0pt,topsep=1pt]
    \item \textbf{Stage-V: Virtual Packet Modality Selection.} Exogenous packets are enqueued into virtual queues indexed by modality; devices then use an MCF solver to balance workloads by switching packets across modalities (detailed in Section~\ref{sec:virtual}).
    \item \textbf{Stage-P: Physical Routing and Scheduling.} Based on the updated virtual queues, SP-BP routing and scheduling are performed independently  and in parallel within each conflict graph layer under conflict-aware transmission constraints (Section~\ref{sec:physical}).
    \item \textbf{Data Transmission:} Packets are transmitted over scheduled links at the allocated rates. 
\end{itemize}

To formally explain these stages, we extend the SP-BP formulation to the multi-layer graph $\ccalG^m$. 
For each subnode $i_m$ and commodity $c$, the biased backlog is defined as
\begin{equation}\label{eq:U_virtual}
    U_{i_m}^{(c)}(t) = q_{i_m}^{(c)}(t) + B_{i_m}^{(c)},
\end{equation}
where $B_{i_m}^{(c)} \geq 0$ is the shortest-path bias from $i_m$ to destination $c$ computed over $\ccalG^m$. 
The pressure of commodity $c$ on any directed edge $e=(v,u)\in\ccalE$ is $U_{vu}^{(c)}(t)=U_v^{(c)}(t)-U_u^{(c)}(t)$, where $e$ may be virtual or physical.

As abstract intra-device transfers, virtual links have no intrinsic rates
to differentiate connectivity and transmission capabilities of receiving modalities.
To break this symmetry, we define the instantaneous rate of each virtual
link \(e^v=(i_n,i_m)\), \(n,m\in\ccalM^+\), as the average instantaneous
rate of the outgoing physical links of the receiving subnode \(i_m\):
\begin{equation}\label{eq:v_rate}
    \grave{\bbr}_{e^v}(t)
    = \frac{1}{|\ccalE_{i_m}^{p,+}|}
    \sum_{e^p\in\ccalE_{i_m}^{p,+}}
    \grave{\bbr}_{e^p}(t),\quad
    \bbr_{e^v}=\mathbb{E}[\grave{\bbr}_{e^v}(t)],
\end{equation}
where \(\ccalE_{i_{m}}^{p,+}\) is the outgoing physical-link set defined in Section~\ref{sec:system}. 
The virtual link rate in \eqref{eq:v_rate} imposes an artificial limit on virtual transmission across modalities.
Each edge \(e\in\ccalE\) is assigned a shortest-path weight: \(\delta_e=\bar r^{p}r_{\max}^{p}/\bbr_e\) for \(e\in\ccalE^p\) and \(\delta_e=1/\bbr_e\) for \(e\in\ccalE^v\), where \(\bar r^{p}\) and \(r_{\max}^{p}\) are the average and maximum long-term physical-link rates.
The bias set \(\{B_{v}^{(c)}\}_{v\in\ccalV,c\in\ccalV^0}\) can be found by all-pairs shortest path (APSP) algorithms~\cite{bernstein2021distributed}  on \(\ccalG^m\) with edge weights \(\{\delta_e\}_{e\in\ccalE}\).

\subsection{Coupled Virtual and Physical Stages}

The MaxWeight formulation of SP-BP is given by~\cite{neely2005dynamic,zhao2024tmlcn,zhao2026ton}
\begin{equation}
\label{eq:P_MW}
\big(\mu^\star(t),\nu^\star(t)\big)=
\argmax_{(\mu,\nu)\in\Pi(t)}\ \Big( W_{\mathrm{p}}^t+W_{\mathrm{v}}^t\Big),
\end{equation}
where \(\Pi(t)\) is the feasible joint action set defined by the link-rate, backlog-availability, 
and conflict-graph constrints, and $ W_{\mathrm{p}} $ and $W_{\mathrm{v}}$ are defined as
\begin{equation}
W_{\mathrm{p}}^t\!=\!\!\sum_{e\in\ccalE^p}\!\sum_{c\in\ccalV^0}\!
\mu_{e}^{(c)}\!(t)\,U_{e}^{(c)}\!(t),\; 
W_{\mathrm{v}}^t\!=\!\!\sum_{e\in\ccalE^v}\!\sum_{c\in\ccalV^0}\!
\nu_{e}^{(c)}\!(t)\,U_{e}^{(c)}\!(t).
\end{equation}
Solving~\eqref{eq:P_MW} is challenging because it couples conflict-free virtual links with conflict-aware physical links. 
Therefore, we approximate it in two stages: \emph{Stage~V} jointly assigns exogenous and endogenous packets to modalities by solving 
\begin{equation}\label{eq:stage_v}
 \nu(t) = \argmax_{\tilde\nu(t)\in\Pi_\nu(t)} \sum_{e\in\ccalE^v}\!\sum_{c\in\ccalV^0}\!
\tilde \nu_{e}^{(c)}\!(t)\,U_{e}^{(c)}\!(t).
\end{equation}
\begin{equation}
\label{eq:q_prime_update}
\hat q^{(c)}_{v}(t)= q^{(c)}_{v}(t) + \Delta\nu_{v}^{(c)}(t) ,\quad \forall v\in\ccalV,c\in\ccalV^0.
\end{equation}
Next, \emph{Stage~P} updates intermediate pressure $\hat U_{e}^{(c)}\!(t)$ with $ \hat q^{(c)}_{v}(t) $ using~\eqref{eq:U_virtual}, and then solves the following MWIS problem for joint routing and scheduling:
\begin{equation}\label{eq:stage_p}
    \mu(t) = \argmax_{\tilde\mu(t)\in\Pi_{\mu}(t)}  \sum_{e\in\ccalE^p}\!\sum_{c\in\ccalV^0}\!
\tilde \mu_{e}^{(c)}\!(t)\,\hat U_{e}^{(c)}\!(t) \;.
\end{equation}

\begin{figure}[t]
    \centering
    \includegraphics[width=0.95\linewidth]{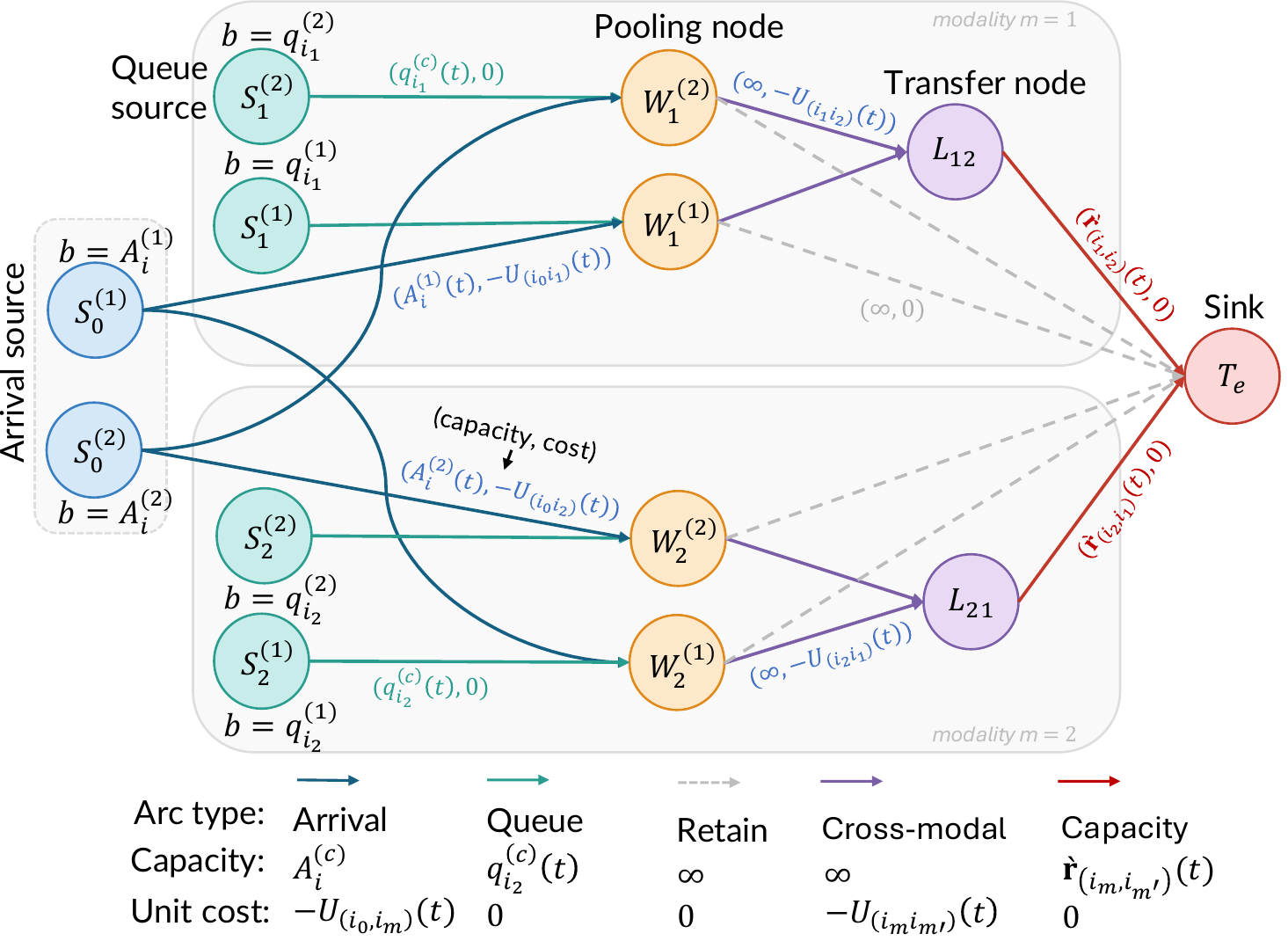}
    \vspace{-0.6em}
    \caption{An exemplary directed acyclic graph $\mathcal{D}_i(t)$ in the MCF formulation of Stage V with $|\ccalV^0|=2$ commodities and $|\ccalM_i|=2$ link modalities.}\label{fig:flow_network}
    \vspace{-1.2em}
\end{figure}

\subsection{Intra-device Virtual Routing}\label{sec:virtual}
Since virtual link sets $\ccalE^v_i$ and $\ccalE^v_j $ are always disjoint for $i\neq j$, \eqref{eq:stage_v} is equivalent to independently solving subproblems on individual devices. 
Because virtual links are conflict-free, we formulate \emph{Stage~V} in~\eqref{eq:stage_v} for device $i$ as an MCF problem on a directed acyclic graph (DAG) \(\mathcal{D}_i(t)\), as illustrated in Fig.~\ref{fig:flow_network}. 
Each virtual commodity queue on modality-node $i_m$ serves as a flow source \(S_m^{(c)}\) that supplies $q^{(c)}_{i_m}(t)$ packets for $m\in\ccalM_i^+,c\in\ccalV^0$. 
A pooling node \(W_m^{(c)}\) is created for each $m\in\ccalM_i, c\in\ccalV^0$, and a transfer node \(L_{m,m'}\) is created for each pair $m,m'\in\ccalM_i$ and $m\neq m'$, to enforce the virtual link rate constraint.
The sources \(S_0^{(c)},S_m^{(c)}\) are connected to \(W_m^{(c)}\), which is subsequently connected to \(L_{m,m'}\). 
The pooling and transfer nodes are all connected to the sink \(T_e\) that absorbs all flows.

The arc costs and capacities are as follows: 
Injection arcs \(S_0^{(c)}\to W_m^{(c)}\) have capacity \(A_i^{(c)}(t)\) and cost \(-U_{i_0 i_m}^{(c)}(t)\). 
Queue arcs \(S_m^{(c)}\to W_m^{(c)}\) have capacity \(q_{i_m}^{(c)}(t)\) and zero cost. 
Retention arcs \(W_m^{(c)}\to T_e\) have zero cost and infinite capacity, while switching arcs \(W_m^{(c)}\to L_{m,m'}\) have cost \(-U_{e^v}^{(c)}(t)\) for \(e^v=(i_m,i_{m'})\). 
Finally, each \(L_{m,m'}\to T_e\) arc has capacity \(\grave{\bbr}_{e^v}(t)\) and zero cost, so that the assignments respect the virtual link rate constraint. 
Thus, minimizing the total flow cost maximizes the objective in~\eqref{eq:stage_v}, subject to backlog availability and the one-switch-per-slot structure.

The network simplex MCF solver~\cite{orlin1997polynomial} then determines injection flows \(\nu_{i_0 i_m}^{(c)}(t)\) for assigning new arrivals to a modality subnode, and switching flows \(\nu_{e^v}^{(c)}(t)\) for assigning queued packets from one modality to another. 

\subsection{Stage P and Data Transmission}\label{sec:physical}
In \textit{Stage P}, SP-BP~\cite{zhao2024tmlcn} is executed in parallel on disjoint layers of the multi-layer connectivity graph $\mathcal{G}^m$ and multi-layer conflict graph $\ccalG^c$.
Specifically, each modality without cross-modality conflicts forms a separate layer, whereas mutually conflicting modalities share a joint layer containing their cross-modality conflict edges.
The procedure uses the intermediate virtual queues \(\{\hat q^{(c)}_{v}(t)\}\) and consists of commodity selection, preliminary rate allocation, MaxWeight link scheduling, and final rate assignment as detailed in~\cite{zhao2024tmlcn}.
The NP-hard MWIS problem on $\widetilde{\ccalG}^c_m$ in MaxWeight scheduling can be approximated by distributed heuristics such as the local greedy scheduler (LGS)~\cite{joo2012local} and GCN-LGS~\cite{zhao2022twc}.

Stages V and P only manipulate virtual queues without moving any packets in the physical queues; data packets are only dequeued for transmission after Stage P (Fig.~\ref{fig:timeline}) based on the final $\{\mu_{v}^{(c)}(t)\}_{v\in\ccalV,c\in\ccalV^0}$, avoiding intra-device looping.

\subsection{Complexity Analysis}\label{sec:complexity}
\begin{table}[bt!]
\centering
\renewcommand{\arraystretch}{1.4}
\caption{Per-Slot Complexity of SP-BP for MWN}
\label{tab:complexity}
\setlength{\tabcolsep}{1pt}
\resizebox{0.98\linewidth}{!}{\begin{tabular}{l c c c}
Operation & Computation & Comm. rounds & Msg. Size  \\ \hline
Stage~V & \(\ccalO(\widehat{M}^3|\ccalV^0|^2\log(\widehat{M}|\ccalV^0|))\) & -- & -- \\ \hline
Routing &  \(\ccalO(|\ccalV^0|D^n)\) & $\ccalO(1)$ & $\ccalO(|\ccalV^0|\widehat{M})$  \\ \hline
Scheduling~\cite{joo2012local} &  \(\ccalO(D^c\log|\ccalE^p|)\) & $\ccalO(\log|\ccalE^p|)$ & $\ccalO(\widehat{M})$  \\ \hline
Bias (APSP)~\cite{bernstein2021distributed} & -- & $\tilde\ccalO(|\ccalV|/\tau_B)$ & $\ccalO(\log|\ccalV^0|)$ \\ \hline
\end{tabular}}
\vspace{-1.2em}
\end{table}

We denote the number of commodities as \(|\ccalV^0|\), the number of communication modalities on device \(i\) as \(M_i=|\ccalM_i|\), with 
$\widehat{M}=\max_{i \in \ccalV^0} \{M_i\}$, and \(D^n,D^c\) as the maximum node degrees on $\ccalG^n,\ccalG^c$. 
Table~\ref{tab:complexity} summarizes the per-slot local complexity under distributed execution, where the network-wide complexity is governed by the maximum local term.
\emph{Stage~V} solves an MCF on a DAG $\ccalD_i(t)$ with \(\ccalO(M_i^2|\ccalV^0|)\) arcs via network simplex, resulting in a time complexity of \(\ccalO(M_i^3|\ccalV^0|^2\log(M_i|\ccalV^0|))\)~\cite{orlin1997polynomial}.
For SP-BP routing, computing biased pressure weights and selecting the best commodity requires \(\ccalO(|\ccalV^0|D^n)\) operations per slot.
In LGS~\cite{joo2012local}, each physical link compares its weight with those of its contention neighbors, leading to a time complexity of \(\ccalO(D^c)\) per round.

Signaling complexity measures the number of distributed control rounds.
SP-BP routing performs one neighbor queue-state exchange per slot, giving \(\ccalO(1)\).
LGS requires distributed coordination among conflicting links and converges in \(\ccalO(\log|\ccalE^p|)\) rounds~\cite{joo2012local}.
The bias matrix is reused until the topology changes.
With recomputation period \(\tau_B\), distributed APSP~\cite{bernstein2021distributed} requires \(\tilde\ccalO(|\ccalV|/\tau_B)\) amortized rounds, where \(\tilde\ccalO(\cdot)\) suppresses polylogarithmic factors.

Message-size complexity measures information per signaling round. 
SP-BP routing sends \(\ccalO(|\ccalV^0|\widehat{M})\) per-modality queue counters per neighbor exchange. 
With differential updates, only changed counters are sent, reducing overhead in practice.
LGS messages carry \(\ccalO(\widehat{M})\) link weights and scheduling states. 
Each APSP round propagates one bias value per commodity, giving \(\ccalO(|\ccalV^0|)\).

\section{Simulation Results}\label{sec:experiments}
We evaluate the proposed multi-layer SP-BP on simulated wireless networks.
For each $|\ccalV^0|\in\{20,30,\ldots,100\}$, we generate 10 independent random topologies in a 2D square with node density $8/\pi$.
We consider three concurrently operating communication modalities (5 GHz, 2.4 GHz, and 900 MHz UHF) over the same physical nodes.
Each modality's communication range induces a distinct connected layer of $\ccalG^m$, and its interference radius determines the modality-specific conflict graph $\widetilde{\ccalG}^c_m$.
Cross-modality interference is absent. The interference radii satisfy $\rho_{\mathrm{5G}}<\rho_{\mathrm{2.4G}}<\rho_{\mathrm{UHF}}$, and larger values yield higher conflict degrees and fewer transmission opportunities.

We assign modality layer $m$ a total rate budget of $26 w_m E_m$, where $E_m$ is its number of undirected edges and $[w_0, w_1, w_2] = [0.5, 0.35, 0.15]$. 
To prevent any modality from dominating spatially, this budget is distributed among its $E_m$ edges as $\bbr_{e}=26E_mw_m\bbp_{e}^{(m)}$, where \( \mathbf{p}^{(m)} \sim \mathrm{Dirichlet}(0.8 \mathbf{1}_{E_m}) \).
Each undirected edge $e$ is then converted into a pair of directed modality-links with equal long-term rates, e.g., $\mathbf{r}_{(i_m,j_m)} = \mathbf{r}_{(j_m,i_m)}=\bbr_e$. 
Although symmetric rates are used for simplicity, our routing scheme fully supports asymmetric cases. 
Finally, to simulate fading, real-time link rates are sampled at each slot as $\grave{\mathbf{r}}_{e^p}(t) \sim \mathcal{N}(\mathbf{r}_{e^p}, 2)$, truncated within $\mathbf{r}_{e^p} \pm 9$.
For each network topology, we evaluate 10 flow configurations, each containing a number of source-destination pairs sampled uniformly between $\lfloor 0.3|\mathcal{V}^0| \rfloor$ and $\lceil 0.5|\mathcal{V}^0| \rceil$. 
Each combination of network topology and flow configuration defines a test instance, which is simulated for $T=1000$ slots under each routing algorithm, yielding 100 test instances per network size and traffic setting.

\begin{figure}[t]
\centering
\includegraphics[width=0.8\linewidth]{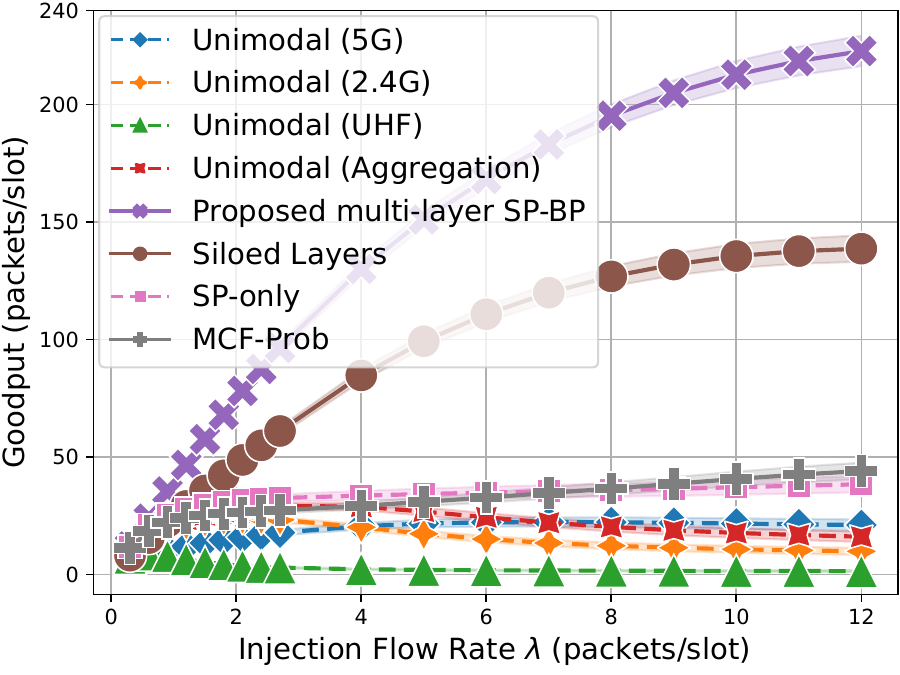}
\vspace{-0.6em}
\caption{Average network goodput (average packets delivered network-wide per slot) versus per-flow injection rate $\lambda$ in 100-node MWNs with streaming traffic; shading denotes 95\% bootstrap confidence intervals for the mean.}
\vspace{-1.2em}
\label{fig:goodput_compare}
\end{figure}

We compare our multi-layer framework with a total of seven baselines grouped in two categories, based on whether SP-BP is employed.
All baselines employ the same local greedy scheduler~\cite{joo2012local} to approximate MWIS scheduling. 

\noindent\textbf{SP-BP operating on baseline graph models:} 
\begin{itemize}[leftmargin=1em,labelsep=0.2em,itemsep=0pt,topsep=1pt]
\item \textbf{Unimodal} (\textit{5G}, \textit{2.4G}, \textit{UHF}): ablation baselines that restrict all network traffic to one specific modality.
\item \textbf{Unimodal} (\textit{Aggregation}): a single-layer graph variant in which the link rate between a node pair is aggregated across all modalities in the original Het-MuNet.
\item \textbf{Siloed layers}: a simplified version of the proposed algorithm, in which each packet is assigned at injection to a single graph layer based on the current biased backlog $U_{i_m}^{(c)}(t)$ and remains on that layer for its entire route.
\end{itemize}
\noindent\textbf{Non-BP routing with decoupled MaxWeight scheduling:}
\begin{itemize}[leftmargin=1em,labelsep=0.2em,itemsep=0pt,topsep=1pt]
\item \textbf{SP-only}: a delay-weighted shortest-path routing policy operating on a collapsed single-layer graph that retains only the highest-rate modality for each physical link.
\item \textbf{MCF-Prob}: a static probabilistic forwarding policy derived by solving an offline multi-commodity MCF problem over the physical
links of our layered topology, where each link capacity is set to be the link rate divided by conflict degree.
\end{itemize}

\subsection{Network Goodput Analysis}
The impact of the routing schemes on network capacity is evaluated under streaming flows with Poisson arrivals at identical rates $\lambda\in[0.3,12]$.
As shown in Fig.~\ref{fig:goodput_compare}, the proposed \textit{multi-layer SP-BP} achieves the highest mean goodput across all evaluated injection rates, with its advantage widening under higher load. 
\textit{Siloed layers} also continues to improve under higher load, but stays below \textit{multi-layer SP-BP} because it lacks dynamic cross-modality switching.
In contrast, \textit{SP-only} and \textit{MCF-Prob} increase much more slowly, whereas the unimodal baselines saturate at substantially lower goodput. 
This saturation occurs under excessive load because the MWIS scheduler prioritizes links near source nodes while starving those near destinations~\cite{zhao2024tmlcn}. 
In particular, \textit{Unimodal (Aggregation)} peaks near \(\lambda=3\) and then gradually declines.

\subsection{Performance under Mixed Traffic}

\begin{figure*}[t!]
\centering

\noindent\makebox[\linewidth][c]{    \includegraphics[width=0.9\linewidth]{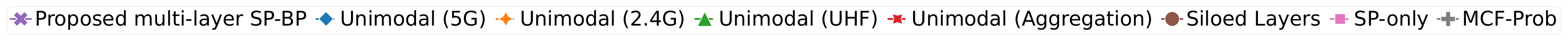}}
\vspace{-1.2em}

\begin{minipage}[t]{0.245\linewidth}
    \centering
    \includegraphics[width=\linewidth]{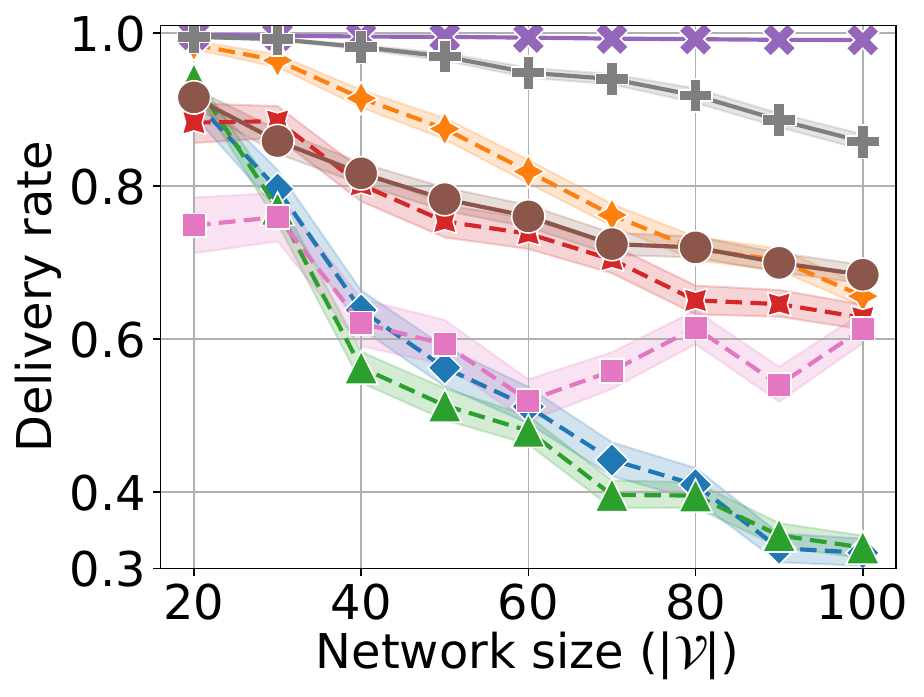}
    \footnotesize (a) Stream delivery
\end{minipage}
\begin{minipage}[t]{0.245\linewidth}
    \centering
    \includegraphics[width=\linewidth]{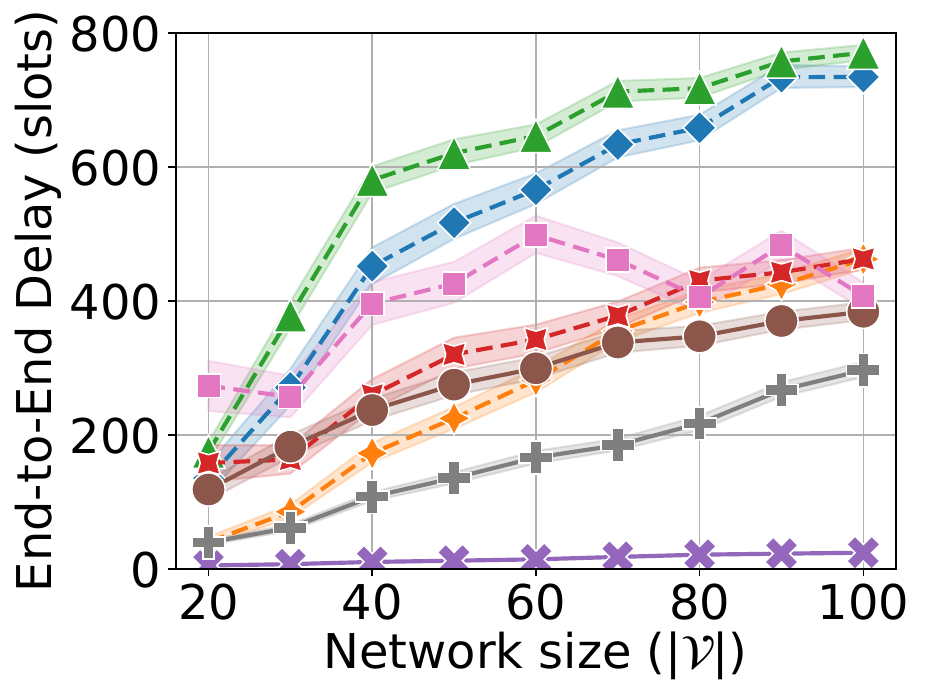}
    \footnotesize (b) Stream delay
\end{minipage}
\begin{minipage}[t]{0.245\linewidth}
    \centering
    \includegraphics[width=\linewidth]{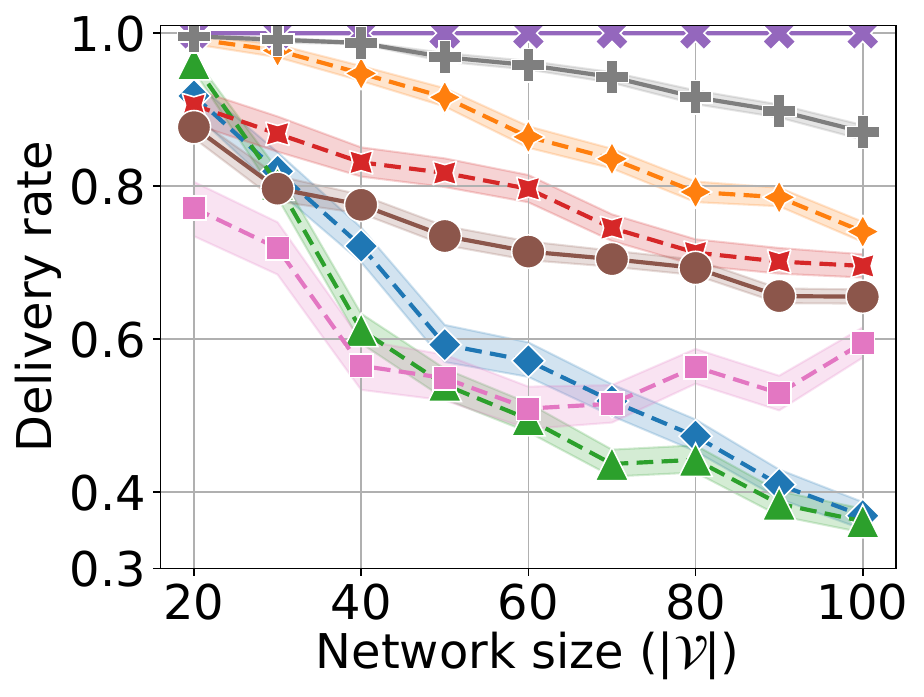}
    \footnotesize (c) Bursty delivery
\end{minipage}
\begin{minipage}[t]{0.245\linewidth}
    \centering
    \includegraphics[width=\linewidth]{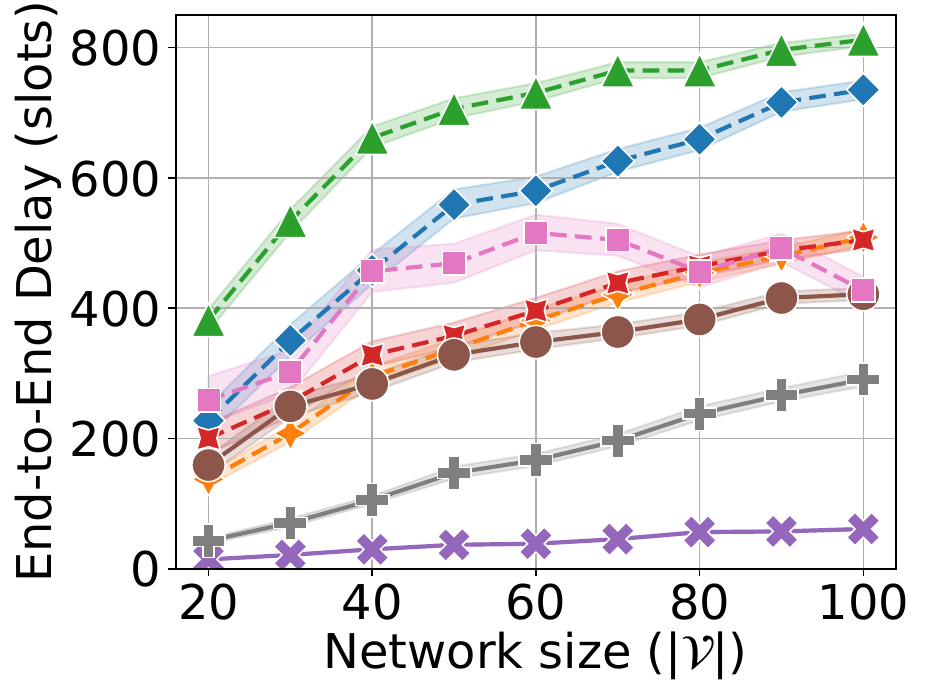}
    \footnotesize (d) Bursty delay
\end{minipage}
\vspace{-0.1in}
\caption{Routing performance under low-to-medium mixed traffic: (a), (c) packet delivery ratio and (b), (d) end-to-end delay for streaming (a), (b) and bursty (c), (d) traffic. Undelivered packets are assigned a delay of \(T\); shading denotes 95\% bootstrap confidence intervals for the mean.}
\label{fig:performance_scecon}
\vspace{-1.2em}
\end{figure*}

In this experiment, the traffic is a mixture of streaming and bursty flows, assigned independently with equal probability. Streaming flows model continuous data injection with Poisson arrivals at a fixed rate $\lambda_s\sim U(0.2,1.0)$, whereas bursty flows model short-lived, high-intensity traffic with Poisson arrivals at rate $\lambda_b\sim U(6.6,33.0)$ for $t<30$ and none thereafter.

Fig.~\ref{fig:performance_scecon} shows that \textit{multi-layer SP-BP} achieves the best tradeoff between delivery and delay under streaming traffic, maintaining delivery rates above $0.99$ across all tested network sizes while yielding the lowest end-to-end delay.
Although \textit{MCF-Prob} closely approaches it in small networks, its static flow split cannot adapt as the network grows, reducing delivery to approximately $0.86$ at $|\ccalV^0|=100$ and forcing longer routes with higher delay.
In contrast, unimodal schemes and \textit{Siloed layers} show a faster decline in delivery and higher delay.
Under bursty traffic, similar scaling trends hold, with the proposed method maintaining robust performance across network sizes.
The relative ordering of \textit{Siloed layers} and \textit{Aggregation} reverses across traffic types. 
\textit{Aggregation} performs better under bursty traffic by pooling modal capacities, whereas \textit{Siloed layers} benefits from parallel transmissions under streaming traffic.

\section{Conclusions}
\label{sec:conclusions}
We developed a fully distributed, multi-layer SP-BP scheme that formally couples dynamic packet-to-interface assignment with packet routing and link scheduling in Het-MuNets. 
In interference-constrained environments, this joint optimization significantly improves resource utilization and manages inter-flow interference, yielding superior goodput, latency, and packet delivery rates compared with static forwarding policies or siloed SP-BP baselines. 
Ultimately, our approach successfully navigates the tightly coupled decision space of Het-MuNets to unlock their full potential. 
Future work will explore advanced bias formulations to further optimize energy efficiency, reduce radio footprints, and enhance resilience to mobility and link failures.

\bibliographystyle{ieeetr}
\bibliography{mylib,refs}

\end{document}